\documentclass[12pt,a4paper,oneside,preprintnumbers]{article}
\usepackage{graphicx,sectsty,amssymb,amsmath}
\usepackage[linkcolor={blue},citecolor={red},colorlinks=true]
{hyperref}
\usepackage[margin=2cm]{geometry}
\usepackage{multirow}
\usepackage{authblk}

\newcommand{\eg}{{\it e.g.}}
\newcommand{\beq}{\begin{equation}}
\newcommand{\eeq}{\end{equation}}

\newcommand{\qmax}{Q_{\rm max}}
\newcommand{\yir}{y_{\mathrm IR}}
\newcommand{\yuv}{y_{\mathrm UV}}

\begin{document}

\title{{\small \begin{flushright} CERN-PH-TH/2012-202 \end{flushright} } \vspace{0.5cm}   
\bf Flavor Beyond the Standard Universe}

\author[1]{\small Gian F. Giudice}
\author[1,2]{\small Gilad Perez}
\author[2]{\small Yotam Soreq}

\affil[1]{\it\small CERN, Theory Division, CH1211 Geneva 23,
Switzerland}

\affil[2]{\it\small Department of Particle Physics and
Astrophysics, Weizmann Institute of Science, \newline
Rehovot 76100, Israel}

\date{}
\maketitle

\begin{abstract}
We explore the possibility that the observed pattern of quark
masses is the consequence of a statistical distribution of
Yukawa couplings within the multiverse. We employ the anthropic
condition that only two ultra light quarks exist, justifying the
observed richness of organic chemistry. Moreover, the mass of
the recently discovered Higgs boson suggests that the top
Yukawa coupling lies near the critical condition where the
electroweak vacuum becomes unstable, leading to a new kind of
flavor puzzle and to a new anthropic condition. We scan Yukawa
couplings according to distributions motivated by high-scale
flavor dynamics and find cases in which our pattern of quark
masses has a plausible probability within the multiverse.
Finally we show that, under some assumptions, these
distributions can significantly ameliorate the runaway behavior
leading to weakless universes.
\end{abstract}

\section{Introduction}

The discovery of the Higgs boson sheds light on the origin of
electroweak (EW) symmetry breaking but leaves open the problem
of why the weak force is so much stronger than the
gravitational force. Despite the enormous experimental and
theoretical effort we are still in the dark. We currently have
no indications for dynamics beyond the Standard Model (SM) and 
this raises the question of whether the weak scale is dynamically
stabilized in nature or not. At present we cannot discard the
possibility that the EW breaking sector is unnatural as a
result of environmental selection effects~\cite{Agrawal:1998xa}. 
The common wisdom is that, due to the vast landscape of configurations
that are local energy minima, string theory does not uniquely
predict the spectrum of particles and interactions as observed
in our universe~\cite{douglaskachru}.  Eternal inflation might
then generate an enormous number of causally-disconnected
``pocket universes,'' each with its own laws of physics~\cite{eternal}.

While the LHC searches are still ongoing, it is too early to
draw conclusions regarding the naturalness or unnaturalness of
the EW breaking sector. The only new piece of data at our
disposal is the mass of the Higgs boson, which has been found
to be about 125--126 GeV~\cite{HiggsResults}. It is interesting
that the preferred Higgs mass and the top Yukawa are close to
their critical values for vacuum
stability~\cite{EliasMiro:2011aa}. This criticality may just be
a mere coincidence, but it is also possible that the special
value of the top Yukawa is the result of some underlying statistics 
that pushes the coupling towards an environmental boundary.

\begin{figure}[tb]
  \centering
  \includegraphics[width=.8\textwidth]{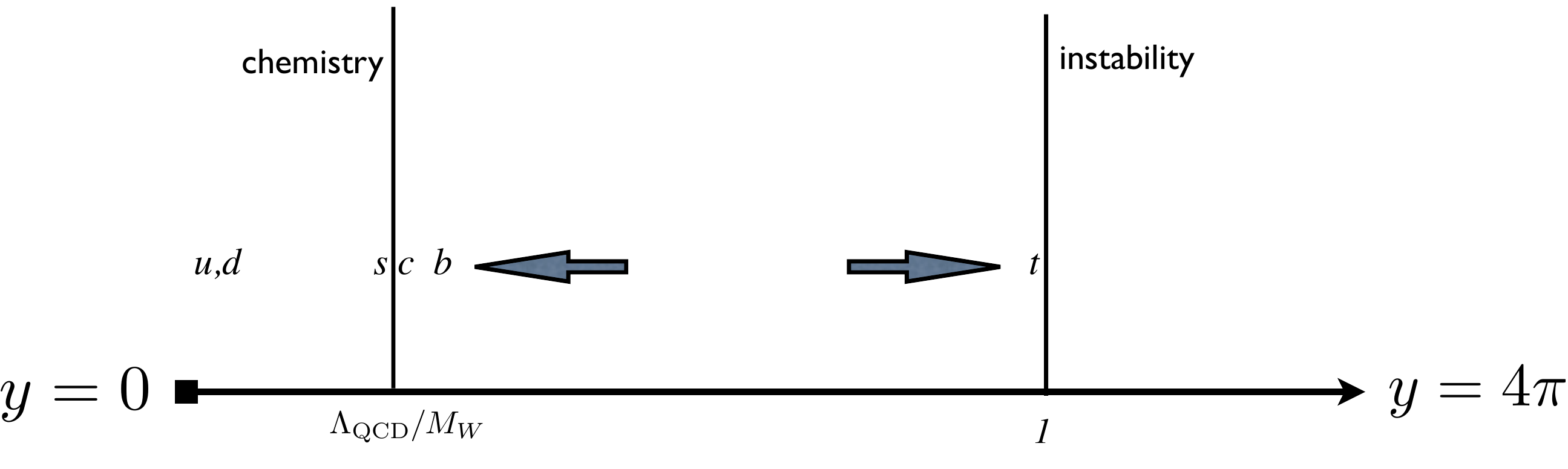}
  \caption{The distributions of quark masses and the anthropic 
boundaries. \label{fig:Anbound}}
\end{figure}

The identification of environmental boundaries for quarks is a
highly non-trivial task from the following two main reasons.
The first is technical in nature: it involves controlling the
way masses, forces, and other physical observables vary when
couplings are scanned. This typically requires mastering
non-perturbative phenomena as well as complicated sets of
coupled equations. The second is more fundamental and is due to
the fact that, even if we are able to fully control the
response to variations in the fundamental laws of nature, one
needs to identify the conditions for which
hospitable universes can exist. Below we shall not attempt to
fully address these two challenges. Instead we shall use a more
minimal and weak criterion where we identify an environmental
boundary with the condition that the structure of matter and
chemistry are not drastically different with respect to our
universe. The analysis of Jenkins et al.~\cite{Jaffe:2008gd}
showed that organic chemistry similar to the one of our own
universe generically requires the presence of exactly two ultra
light quarks, with masses well below $\Lambda_{\rm QCD}$. We 
can view vacuum stability as a new anthropic condition. Assuming a single heavy 
flavor and a fixed  Higgs mass of $125\,$GeV, vacuum metastability requires the
corresponding Yukawa to be below roughly $y_t \leq 1.03$ (or
$m_t\leq 179\,$GeV). The observed spectrum and the above
anthropic boundaries are illustrated in Fig.~\ref{fig:Anbound}.
It is interesting to note that the distribution of masses shows a
pattern, beyond just being hierarchical, in the sense that the
four heavier quarks are distributed close to the boundaries,
namely, $s,c,b$ are close to the ``organic chemistry" boundary,
while the top is close the the electroweak instability
boundary. No states are found in the ``limbo" related to an
intermediate mass range. The LHC result on the Higgs mass is an 
important input because
it suggests that the top Yukawa lies close to a catastrophic
situation. This criticality can be considered as related to a
new problem, {\it the top anthropic flavor puzzle}. 

In this paper, we will address the question of whether the
observed flavor structure can be understood in terms of
parameter scanning with anthropic conditions. For simplicity,
we will focus only on the hierarchies in the spectrum of
quarks. The pattern of CKM mixing angles can be understood as a
consequence of such hierarchies in masses, and we will not
discuss it here. Similarly, we will not address the issue of lepton 
masses, although some of our considerations about quarks can be
extended to the charged lepton sector as well. Our goal is not
to explain the fine details of the quark mass spectrum, but
rather to investigate possible mechanisms that explain its
general structure. We will show that statistical
interpretations of the mass spectrum can exist. Namely, if the
distribution of Yukawa couplings in the multiverse is such that
it rises towards both boundaries (corresponding to the directions of 
the two arrows in Fig.~\ref{fig:Anbound}) then the observed pattern can be
explained. Note that in some sense this is more challenging
than explaining the cosmological constant, as in that case one
requires a monotonic distribution that peaks towards larger
values. In the case presented by us (and illustrated in
Fig.~\ref{fig:Anbound}) a two-peak distribution is required.
Motivated by the observed flavor hierarchies we shall consider
ans\"atze for the Yukawa couplings that capture the current
wisdom regarding generating hierarchies. We discuss various
effects that generate not only preference towards small Yukawa
but also lead to a second peak of the distribution for large
Yukawa.

So far we have discussed distributions in Yukawa couplings, but 
did not consider the implications of scanning over
the Higgs VEV and the Higgs mass. In principle,
one can argue that the parameters of the Higgs potential are 
related to electroweak dynamics and the question of their origin is completely 
orthogonal to the discussion related to the nature of the observed flavor sector.
Thus in the main part of this work we shall just hold the the
Higgs parameters (quartic coupling and VEV) to their current
values. However, there is an important reasons to go beyond
this assumption (see also \cite{Feldstein:2006ce}).
In cases where the Yukawa hierarchies are generated by dynamics (say at the
Planck scale), one generically expects a runaway behavior
towards universes with large Higgs VEV and very small
Yukawas~\cite{Gedalia:2010iy}. Such possibility, called ``the
weakless universe"~\cite{Harnik:2006vj}, is perfectly viable
from an anthropic point of view. We will show that, under
certain assumptions, the same toy example that accounts for the
observed quark mass spectrum can also greatly ameliorate or
even eliminate the runaway behavior towards the weakless
universe.

\section{Flavor Dynamics} \label{sec:fd}

Various mechanisms that address the flavor puzzle have been
studied in the literature~\cite{froggatt-nielsen,
strong-dynamics,PartialCcomp,split-fermions}, yet all of them can 
be summarized by a simple formula for the effective Yukawa
couplings $y$:
\beq \label{eq:epsQ}
y \propto  \epsilon ^Q \,,
\eeq
where $\epsilon$ is some small parameter and $Q$ is a flavor-dependent charge. We assume
that flavor dynamics occur at some high energy scale.

We can broadly distinguish between two classes, which have
similar parametric dependence but lead to very different
behavior. (i) The first class includes Froggatt-Nielsen models
with horizontal U(1)~\cite{froggatt-nielsen}, as well as split
fermions in flat extra dimensions~\cite{split-fermions,
Kaplan-Tait}. For Froggatt-Nielsen, the parameter $\epsilon$ is
matched to the ratio between the flavon VEV and the fundamental
scale and $Q$ to the absolute value of the corresponding
charge. For split fermions $\epsilon$ can be identified roughly
with $\exp(-R/d)^{n}$ ($R$ being the extra dimension size and
$d$ the width of the fermion wave function, with $n=1,2$ for a
constant or linear bulk mass respectively). The parameter $Q$
corresponds to $(\Delta x_5/R)^n$, where $\Delta x_5$ is the
separation between the fermion localization along the extra
dimension, $x_5$. Within this class of models, $Q$ can only
take non-negative values. (ii) The second class includes models
with strong dynamics or models within the warped extra
dimension framework, where $\epsilon$ can be identified roughly
with $(\Lambda_{\rm IR}/\Lambda_{\rm UV})$ and $Q$ is the
anomalous dimension~\cite{PartialCcomp,NelsonStrassler,RS}. 
The peculiarity of this class is that $Q$
can be both positive and negative. To distinguish between the
two cases we use a different convention for the exponent of
this class, and we denote it by $\gamma$.

Starting from class (i), we assume that the parameters $Q$
and $\epsilon$ scan over different vacua within the multiverse.
Here we assume that their probability distribution functions
(PDF) can be described by general power laws
\beq
\label{Q_eps_dist}
p_Q(Q) \propto Q^n \,,  \qquad
p_\epsilon(\epsilon) \propto \epsilon^m \, ,
\eeq
where $Q\in\left[0,\qmax \right]$ and
$\epsilon\in\left[0,1\right]$. Since this mechanism is used to
explain the existence of light quarks, we
consider $p_Q(Q)$ favoring larger $Q$ values, hence
$n>0$. The resulting Yukawa PDF for $y\in\left[0,y_0 \right]$
is calculated by performing the integral:
\beq
p^{\rm (i)}_y(y)=\int_0^{\qmax} dQ \int_0^1 d \epsilon\, \,
p_Q(Q) \, p_\epsilon(\epsilon) \,\delta \left(y-y_0 \epsilon^Q
\right) \, ,
\eeq
which leads to
\beq \label{yuk_dist_full}
p^{\rm (i)}_y(y)\propto 
\frac1y E_{1+n}\left[ -\frac{1+m}{\qmax} \log(y/y_0)\right] \,,
\eeq
with $m>-1$ and $E_\nu(x)$ being the exponential integral
function, and where the normalization is yet to be fixed. Note
that $\qmax$ and $m$ appear together in the form
$(1+m)/\qmax\,$, so that in practice they are not independent
parameters, and we set $m=2$. A natural choice for $y_0$ would
be the maximal perturbative value of $4\pi$.

An interesting limit is obtained by taking $\qmax$ to be very
large (yet not strictly $\qmax \to \infty$). For a positive $n$
in Eq.~\eqref{Q_eps_dist}, this means that $Q$ would be mostly
concentrated close to $\qmax\,$. We can thus approximate this
limit by plugging $\delta(Q-\qmax)$ into the calculation of
$p^{\rm (i)}_y(y)$, which gives
\beq \label{yuk_dist_limit}
p^{\rm (i)}_y(y)  \rightarrow \left( \frac{y}{y_0}
\right)^{\frac{1+m}{\qmax}-1} \sim \frac1y  \,.
\eeq
Consequently, the rather complicated function in
Eq.~\eqref{yuk_dist_full} behaves asymptotically as a simple
scale-invariant distribution.

Next we discuss the second class of flavor models, where $y=y_0 
\epsilon^\gamma$ and now the exponent $\gamma$ (replacing $Q$) can take also negative
values. This allows us to obtain heavy quarks with $y>y_0\,$,
which means that $y_0$ should be taken smaller than $4\pi$. This 
class can lead to a richer structure of Yukawa distributions.  If
we adopt a power law distribution for $\gamma$ as in
Eq.~\eqref{Q_eps_dist} for $Q$, we will get a PDF for the
Yukawa similar to that in Eq.~\eqref{yuk_dist_full}, but with
an absolute value on the argument of the exponential integral
function, which is still a monotonically decreasing function of
the UV Yukawa. However, going back one step and fixing
$\epsilon$, the behavior of the resulting PDF changes
significantly, as a sharp minimum appears at $y=y_0\,$. In this 
case the function can lead to statistical ``pressure" towards both 
small and large Yukawa couplings. This
can be understood by noting that the PDF coming from the
distribution of $\gamma$ only is $\sim \log^n(y/y_0)/y$,
leading to this minimum. An additional piece
$\log^{-n-1}\epsilon$ appears when $\epsilon$ is also
integrated over. This creates a strong preference towards
$\epsilon=1$, which enhances the probability to have $y \simeq
y_0$ thus washing away the minimum. The minimum is not 
washed away when the PDF for $\epsilon$ peaks strongly away 
from unity, for example when there is
preference towards small values, \eg\ $p_\epsilon(\epsilon) \sim
\exp (-\alpha \, \epsilon)$ with a large enough $\alpha$.

In practice, for class~(ii) models we choose a 
distribution for $\gamma$ which favors large
(positive or negative) values much more pronouncedly than the
power law PDF, and thus leads to either light or heavy quarks:
\beq \label{crazy}
p_\gamma(\gamma) \propto e^{a \gamma^2} \,,
\eeq
with $a>0$. For simplicity, we keep $\epsilon$ fixed in this
case. The resulting (not normalized) Yukawa distribution is
\beq \label{yuk_dist2}
p^{\rm (ii)}_y(y)  \propto \frac1y
\exp\left[a\left(\frac{\log\left( y/y_0\right)}{\log\epsilon}
\right)^2 \right] \, .
\eeq

\begin{figure}[tb]
  \centering
  \includegraphics[width=.45\textwidth]{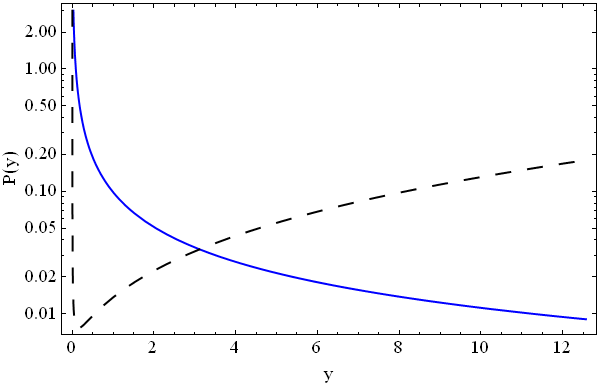} \hspace{1cm}
  \includegraphics[width=.45\textwidth]{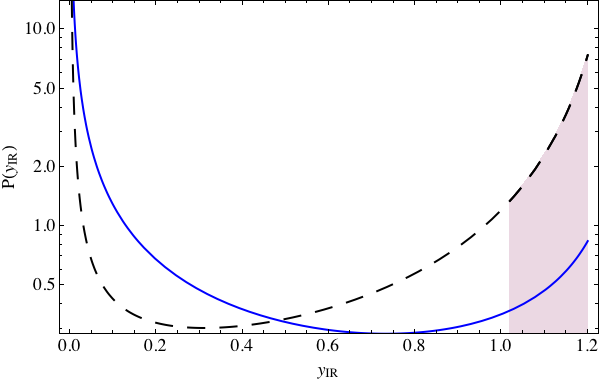}
  \caption{PDF of a single Yukawa coupling. The solid blue  and 
dashed black lines correspond, respectively, to
  the distributions of $p_y^{\rm (i)}$ (Eq.~\eqref{yuk_dist_full}  with 
$n=m=2$ and $\qmax=100$) and $p_y^{\rm (ii)}$ (Eq.~\eqref{yuk_dist2} with 
$a=3.1$, $y_0=6.8 \times 10^{-3}$ and $\epsilon=0.012$). Left 
panel: the original distributions as applied at the Planck scale. Right panel: the
 distributions after one-loop RGE running down to the weak scale. The shaded region is excluded by the requirement of vacuum metastability. \label{fig:1yukpdf}}
\end{figure}

In the left panel of Fig.~\ref{fig:1yukpdf} we show the two
distributions of Eqs.~\eqref{yuk_dist_full} (with $n=m=2$ and
$\qmax=100$) and~\eqref{yuk_dist2} (with $a=3.1$, $y_0=6.8
\times 10^{-3}$ and $\epsilon=0.012$). When $\gamma$ is 
scanned above roughly 1 there is no sensitivity to the range in which
it is varied.

\section{Renormalization Group Effects and the Stability Bound} \label{sec:rge}

We discuss here how inclusion of the running of Yukawa
couplings from the high flavor mediation scale, $\mu_{\mathrm UV}$,
to the electroweak one, tend to favor heavy quarks. This can be
understood because of the presence of a pseudo fixed point at low
scale $\mu_{\mathrm IR}$. Another aspect that affects the range of
possible Yukawa eigenvalues is the requirement that the Higgs
potential is not unstable. In this section we consider the effects of
the renormalization group equations (RGE) and the stability bound on the Yukawa distribution of a
single quark (the case of several flavors is discussed in the following).

\begin{figure}[tb]
  \centering
  \includegraphics[width=.5\textwidth]{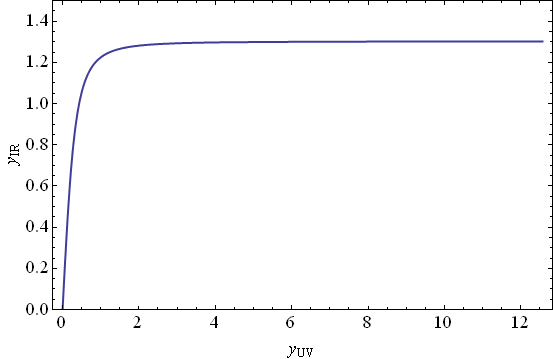}
  \caption{The top Yukawa coupling at the weak scale as a
  function of the initial value at the Planck scale,
  after one-loop RGE running. \label{fig:1yukrge}}
\end{figure}

Taking an initial value for the top Yukawa coupling in the
range $[0,4\pi]$ at the Planck scale, its IR value at the weak
scale is presented in Fig.~\ref{fig:1yukrge}, using one-loop
RGE. It is evident that for any initial value above $\sim$2 at
the Planck scale, the IR Yukawa is 1.3. This can be described
analytically:
\beq \label{iruv}
\yir= \alpha \left( \beta+\yuv^{-2} \right)^{-1/2} \,,
\eeq
\begin{eqnarray}
\alpha(\mu_{\mathrm IR}) &=& \left[ \frac{g_3(\mu_{\mathrm IR})}
{g_3(\mu_{\mathrm UV})}\right]^{8/7}
\left[ \frac{g_2(\mu_{\mathrm IR})}{g_2(\mu_{\mathrm UV})}
\right]^{27/38}
\left[ \frac{g_1(\mu_{\mathrm IR})}{g_1(\mu_{\mathrm UV})}
\right]^{-17/82}\, ,\\
\beta &=& \frac{9}{16\pi^2}\int_{\mu_{\mathrm IR}}^{\mu_{\mathrm 
UV}}\frac{d\mu}{\mu}~\alpha^2(\mu) ~,
\end{eqnarray}
where $g_{1,2,3}$ are the gauge couplings. Since $\beta>1$, the
IR Yukawa quickly reaches an asymptotic value
$\yir=\alpha/\sqrt{\beta}$ for large $\yuv\,$, as evident in
Fig.~\ref{fig:1yukrge}. For $\mu_{\mathrm UV}$ equal to the Planck 
mass, we find
$\alpha=3.6$ and $\beta=7.7$.

If we now apply the RGE result of Eq.~\eqref{iruv} to the
Yukawa distributions in Eqs.~\eqref{yuk_dist_full}
and~\eqref{yuk_dist2} for the initial UV value, we can derive a
PDF for the IR value:
\beq
\begin{split}
p^{\rm (i)}_y(\yir)&\propto \frac{\alpha^2}{\yir
\left(\yir^2\beta^2-\alpha^2\right)} E_{1+n} \left[
-\frac{1+m}{\qmax} \log \left( \frac{\yir/y_0}{\sqrt{\alpha^2-
\yir^2 \beta^2}} \right)
\right]\,,\\
p^{\rm (ii)}_y(\yir)& \propto \frac{\alpha^2}{\yir
\left(\yir^2\beta^2-\alpha^2\right)} \exp\left[a \log^2 \left(
\frac{\yir/y_0}{\sqrt{\alpha^2- \yir^2 \beta^2}}
\right)/\log^2\epsilon \right] \,.
\end{split}
\eeq
These are shown on the right hand side of Fig.~\ref{fig:1yukpdf}.
Interestingly, the IR Yukawa assumes a binary-like
distribution, in which either small or large values are
favored, while intermediate values are not.

The stability of the Higgs potential has been thoroughly
studied in the literature (see {\it e.g.}~\cite{EliasMiro:2011aa,stability,
Casas:1994qy,Isidori:2001bm}). The top Yukawa
drives the Higgs quartic towards small values and as a result
the vacuum becomes unstable or at least metastable in a
significant part of the parameter space described by the top
and Higgs masses and the strong coupling.
The stability and metastability bounds for the top Yukawa are given
by~\cite{EliasMiro:2011aa}
\beq \label{stability_condition}
y_t < 0.98+0.0029 \left(\frac{m_h-125.5\,{\rm
GeV}}{1\,{\rm GeV}}\right) +0.0014 \left(\frac{\alpha_s(m_Z)- 
0.1184}{0.0007}
\right) \pm 0.0029
 \,,
\eeq
\beq \label{metastability_condition}
y_t < 1.02+0.0018 \left(\frac{m_h-125.5\,{\rm
GeV}}{1\,{\rm GeV}}\right) +0.0017 \left(\frac{\alpha_s(m_Z)- 
0.1184}{0.0007}
\right) \pm 0.0055 \,,
\eeq
respectively. Assuming no new physics up to the Planck scale,
the conclusion is that present measurements favor the
possibility that our electroweak vacuum is metastable. It is
quite remarkable to note that for a fixed $125$\,GeV Higgs mass
the top Yukawa is less than $3\%$ from making our electroweak
vacuum unstable! Below we use the requirement of metastability
of the Higgs potential as an anthropic upper bound on quark
masses.

\section{Multi-Flavor Analysis and The Quark Spectrum} \label{sec:masses}

We now combine the ingredients above and study the resulting
quark mass structure within a multiverse framework. The
analysis is based on the following assumptions:
\begin{itemize}
\item There is no new physics beyond the SM up to the
    Planck scale. (As a sensitivity check, we have also
    analyzed lower UV scales; as long as the flavor
    mediation scale is large enough, the qualitative
    behavior described below is unchanged.)
\item The Yukawa distributions of
    Eqs.~\eqref{yuk_dist_full} and~\eqref{yuk_dist2} are
    therefore applied at the Planck scale, so that the
    associated flavor dynamics takes place at that scale or
    above. Alternatively, we examine also the case where
    the Yukawa distribution applies at the weak scale, and
    the RGE play no role.
\item The existence of the lightest two (and only two)
    quarks is ensured by anthropic arguments.
\item Only the Yukawa sector of the SM is being scanned
    over the multiverse, while the gauge and Higgs
     parameters (quartic coupling and bilinear term) are
     held fixed. The implications of scanning over the
     Higgs mass are discussed below in
     section~\ref{VEVscan}.
\end{itemize}
The Yukawa couplings of the four heaviest quarks are generated
(uncorrelated) at the Planck scale, and then evolved down to
the weak scale (equal to the Z mass) using one-loop RGE.
In order to represent the quark mass pattern, or more precisely
the large difference between the heavy top quark and the light
strange, charm and bottom quarks, we define two mass regions:
``light'' quarks for which the weak scale Yukawa eigenvalues
reside in the range corresponding to roughly 60~MeV up to 
10~GeV, and
``heavy'' quarks above 90~GeV (Yukawa coupling greater than
1/2), all evaluated at the Z mass.

On top of the above, we consider the effect of the
metastability bound as another anthropic requirement on the
quark mass distribution. Note that we cannot simply use
Eq.~\eqref{metastability_condition} above, since it only
applies for one heavy quark. Instead, we use the (simplified)
condition~\cite{Casas:1994qy,Isidori:2001bm,Coleman:1977py}
\beq\label{meta_cond_code}
\lambda(M_{\rm Pl}) > -0.095\, ,
\eeq
where $\lambda$ is the Higgs quartic coupling. In principle,
this should be taken to hold for any $\mu$ between the weak
scale and the Planck scale, and indeed there is always a
minimum of $\lambda(\mu)$ at a scale lower than $M_{\rm Pl}\,$.
However, in our numerical calculations we have verified that if we
evaluate $\lambda$ only at the Planck scale, the result is
hardly affected compared to using the real minimum.

In order to put the numerical results below in proper
perspective, we can assume that there is no correlation between
the PDF for the various quarks, and that the total PDF is simply
the product of the independent individual distributions. In
such a case, it is easy to verify that the optimal single quark
PDF for explaining the observed spectrum with three light and
one heavy quark should be such that
\beq \label{simple_probs}
P_{\rm light}= 75\%\,, \ \,P_{\rm heavy}=25\%\,,
\eeq
and obviously nothing in the region between the two ranges.
Consequently, the probabilities to obtain 3 light quarks and
one heavy quark ($P_{\rm 3l|1h}$), 4 light quarks ($P_{\rm
4l}$), and all other cases ($P_{\rm other}$) are
\beq
P_{\rm 3l|1h}= 42\%\,, \ \,P_{\rm 4l}\simeq 32\%\,, \ \,P_{\rm other}\simeq26\% \,. \label{ideal}
\eeq
We should point out that both the RGE effect as well as the
metastability bound induce correlations between multi-quark
PDFs, and thus the above result should be only viewed as an
approximation for the correct result. However, Eq.~\eqref{ideal} can 
be
used as a point of reference for comparison to the results
below.

We first compute the probabilities for a single light or
heavy quark with the PDFs under consideration, taking into
account the RGE from the Planck scale and the metastability
bound. These are normalized to the probability for a quark
above 60~MeV (as evaluated at the Z mass) and below the
metastability bound. The results are:
\beq
P_{\rm light}^{\rm (i)}=56\% \,, \ \ P_{\rm heavy}^{\rm (i)}=14\% \,,
\qquad P_{\rm light}^{\rm (ii)}=58\% \,, \ \ P_{\rm heavy}^{\rm
(ii)}=26\% \,,
\eeq
where we used $n=m=2$ and $\qmax=100$ for the class~(i)
distribution of Eq.~\eqref{yuk_dist_full} and $a=3.1$, $y_0=6.8
\times 10^{-3}$ and $\epsilon=0.012$ for the class~(ii)
distribution of Eq.~\eqref{yuk_dist2}. Interestingly, these numbers
are not very far from the optimal assignment of
Eq.~\eqref{simple_probs}. As expected, the numbers for class (ii) 
are 
closer to the ideal distribution (especially for $P_{\rm heavy}$), as 
a result of the two-peak distribution. Furthermore, as we discuss 
more extensively below, it pushes the Yukawa close to either the 
small or large anthropic boundaries, consistently with the observed 
spectrum. In particular, this mechanism can account for the new top 
anthropic flavor puzzle.

Next, we perform the following exercise. We scan over UV Yukawa
distributions and compute the probabilities to have various
combinations of light and heavy quarks. The normalization is as
before, which means that we account for the ``chemistry"
requirement in such a way that the full set of anthropically
allowed universes corresponds to a probability of 100\%. We
first consider the class (i) distribution of UV Yukawa
couplings given by Eq.~\eqref{yuk_dist_full} for $n=m=2$ and
taken as a function of $\qmax\,$\footnote{Recall that $\qmax$
enters through the combination $(1+m)/\qmax\,$, so that a large
value for $\qmax$ can be replaced with a value for $m$ close to
-1.}. The results are presented in Table~\ref{tab:ProbAll1}. We
verified that the results do not depend strongly on $n$ nor on
the actual UV scale (that is, if it is much lower than the
Planck scale). Similarly,  the results for the class (ii)
distribution of Eq.~\eqref{yuk_dist2} using the parameters
$a=3.1$, $y_0=6.8 \times 10^{-3}$ and $\epsilon=0.012$ are also
given in Table~\ref{tab:ProbAll1}. In this case, lowering the
UV scale does have some effect on the results below (for
instance, the probability to have one heavy quark and three
light ones goes down from 18\% to 9\% for a UV scale of
$10^{10}$~GeV).

\begin{table}[tb]
\begin{center}
\begin{footnotesize}
\begin{tabular}{||l|l||c|c|c||c|c||}
\hline  \hline
\multicolumn{2}{||c||}{}&\multicolumn{3}{|c||}{class (i), Eq.~
\eqref{yuk_dist_full}} &\multicolumn{2}{|c||}{class (ii), Eq.~
\eqref{yuk_dist2}}  \\
\hline
\multicolumn{2}{||c||}{Probability for}  &$\qmax=10$& $\qmax=100$ 
& No RGE& RGE& No RGE \\
\hline \hline \multirow{2}{*} {0 heavy, 4 light}
&without 	 &0 & 2.4\,\% & 3.2\,\%& 0 & 0 \\
&with   	 &0.44\,\% & 9.7\,\% & -- & 14\,\% & -- \\
\hline \multirow{2}{*} {1 heavy, 3 light}
&without 	 & 0.40\,\% & 13\,\% & 11\,\% & 0 & 0 \\
&with   	 & 2.6\,\% & 9.8\,\% & -- & 18\,\% & -- \\
\hline \multirow{2}{*} {2 heavy, 2 light}
&without 	 & 6.3\,\% & 15\,\% & 14\,\%& 0 &0\\
&with   	 & 3.8\,\% & 2.8\,\% & -- &9.4\,\% & -- \\
\hline \multirow{2}{*} {3 heavy, 1 light}
&without 	 & 15\,\% & 4.4\,\% & 8.2\,\%& 1.0\,\%&0.92\,\%\\
&with   	 & 1.8\,\% & 0.27\,\% & -- &2.2\,\% & -- \\
\hline \multirow{2}{*} {4 heavy, 0 light}
&without   	 & 8.3\,\% & 0.35\,\% & 1.8\,\% & 72\,\%&97\,\%\\
&with   	 & 0.21\,\% & 0 &  --  &0.21\,\% &  --  \\
\hline \multirow{2}{*} {2 or more intermediate}
 &without 	 & 29\,\% & 24\,\% & 20\,\%&1.4\,\% &0\\
&with   	 & 61\,\% & 39\,\% & -- &17\,\% & -- \\
\hline \hline
\end{tabular}
\end{footnotesize}
\end{center}
\caption{Probabilities of different combinations of heavy and
light quarks for class~(i) and~(ii) distributions without
and with the metastability bound (probabilities smaller than
0.1\% were rounded down to zero). For class~(i) we use $n=m=2$
while for class~(ii) $a=3.1$, $y_0=6.8 \times 10^{-3}$ and
$\epsilon=0.012$. The No RGE columns stand for the case without
RGE, applying the Yukawa distributions at the weak scale
(without the metastability bound) and using $\qmax=100$ for
class~(i).} \label{tab:ProbAll1}
\end{table}

From the results presented in the table we can draw the
conclusion that the observed spectrum with one heavy and three
light quarks is plausible in the context of a multiverse in
which Yukawa couplings scan. The RGE have two effects: (i) 
increasing the plausibility to have large Yukawa values, as can be 
seen from Fig.~\ref{fig:1yukpdf}, Eq.~\eqref{iruv} and below; (ii) decreasing 
the probability to have several heavy quarks due to correlations 
among them. 
Specifically, the running of a Yukawa coupling in the presence of 
additional heavy flavors pushes it towards smaller values in the IR.  
 The metastability constraint\footnote{Note that in the case without
RGE, when the PDFs are applied at the weak scale, there is no
consistent way to estimate the metastability bound without
assumptions on the UV completion.} plays a crucial role in 
further reducing the plausibility of 
cases with several heavy quarks, especially for
the class~(ii) models, where the preference for large Yukawas
is much stronger (see Fig.~\ref{fig:1yukpdf}). This also means
that for class~(i), the mass of the top quark is distributed
quite uniformly in the region defined as heavy (that is, above
90~GeV), and we find no explanation for the top Yukawa
anthropic puzzle, {\it i.e.} why $y_t$ is just a few percents
below the boundary. On the other hand, for class~(ii) the
probability to obtain a single heavy quark with Yukawa in the
range $0.75-1$ is twice as large than the probability to obtain
a heavy quark with Yukawa in the range $0.5-0.75$, which
constitutes a better explanation for the above puzzle.

To conclude, we find that, for favorable parameters of the
different distributions, universes with a quark pattern similar
to the one we observe can have probabilities in the range
10-20\%. This has to be compared with the reference case of Eq.~\eqref{ideal}, which gives $P_{\rm 3l|1h}= 42\%$. 

\section{Amelioration of the Runaway Behavior} \label{VEVscan}

In~\cite{Gedalia:2010iy} it was shown that applying a simple
Yukawa distribution as in Eqs.~\eqref{eq:epsQ}
and~\eqref{Q_eps_dist} and using the anthropic requirement for
two light quarks makes habitable universes without weak
interaction to be far more plausible than ours. This comes from
the fact that these distributions allow for arbitrarily low
Yukawa values and this is statistically favored since a
quadratic PDF for the Higgs VEV, $p_v(v) \sim v^2/M_{\rm
Pl}^2\,$, prefers large values of $v$.

This runaway behavior can be cured in models where $y=y_0e^\gamma$, with $\gamma$ scanning according to Eq.~\eqref{crazy}, by imposing the following two 
ad-hoc rules. 
(i)  There is a lower cutoff on the Yukawa, $y_{\mathrm {min}}$. 
Notice that unlike the analysis in \cite{Donoghue:2009me}, where $y_{\mathrm 
{min}}$ is of the order of the electron Yukawa, here we use 
$y_{\mathrm {min}}\ll y_e\,$, such that the precise value of 
$y_{\mathrm {min}}$ has no anthropic significance (see also 
\cite{Hall:2007zj}).   
(ii) $y_0$ scales inversely with the Higgs VEV, such that the part of 
the distribution responsible for light quarks always falls close to the 
upper anthropic bound on the two light quarks (here the term 
``light'' refers to the two lightest quarks required by anthropic 
arguments). Realizing such a rule from first principles requires a 
dynamical mechanism that relates $y_0$ to $\Lambda_{\rm QCD}/
v$.    

\begin{figure}[t]
  \centering
  \includegraphics[width=.6\textwidth]{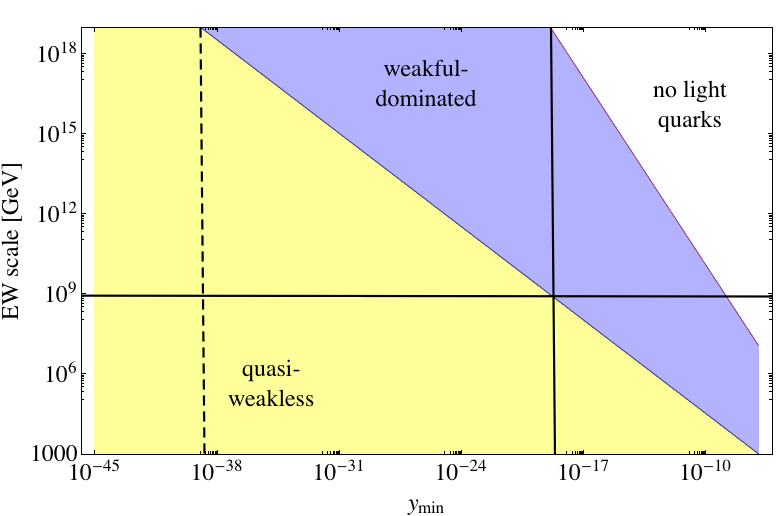}
  \caption{Comparison between the probabilities for the existence of a quark with mass less than $\Lambda_{\rm QCD}$ in a universe with a weak scale 
similar to ours (``weakful") and in
  universes with higher Higgs VEV (``weakless").  The result is shown as a function of the minimum Yukawa coupling $y_{\mathrm 
{min}}$ and the EW scale $v$. In the yellow shaded region both types of universes have a 
$\sim100\%$ probability to produce light quarks, but $R\gg1$ (defined in Eq.~\eqref{eq:R}) and thus  
weakless universes are far more probable. In the blue shaded region we find $R\ll1$, since the 
probability for light quarks drops significantly in the weakless universes, making the weakful 
universe favored. In the white region there are no light quarks in either type of universe. 
\label{fig:runaway}}
\end{figure}

To illustrate the manner in which this mechanism can cure the runway behavior toward  
weakless universes, we consider the following probability ratio 
\beq \label{eq:R}
R = \frac{P\left( m_{\mathrm {light}}\left| v, y_{\mathrm {min}} \right. 
\right)}{P\left( m_{\mathrm {light}}\left| 174\,{\rm GeV}, y_{\mathrm {min}} \right. 
\right)}\times \frac{p_v(v)}{p_v(174\,{\rm GeV})}
\, .
\eeq
Here $p_v(v)\sim v^2$ is the PDF for the Higgs VEV and $P( m_{\mathrm {light}}| v, y_{\mathrm {min}})$ is the probability to obtain a quark mass below $\Lambda_{\rm QCD}$, as we scan $\gamma$ according to Eq.~\eqref{crazy}, for fixed values of the Higgs VEV $v$ and $y_{\mathrm {min}}$.
Thus, $R\ll1$ means that the ``weakful" universe (where the weak 
scale matches our own)  is favored over universes with a higher 
Higgs VEV equal to $v$. By adopting the same values for the
parameters as before with the rescaling of $y_0$
($a=3.1$ $\epsilon=0.012$, and $y_0=6.8
\times 10^{-3}\times 174\,{\rm GeV}/v$), we find that for a rather 
wide range of $y_{\mathrm {min}}$ the weakful universe is favored, 
as shown in the blue shaded region of Fig.~\ref{fig:runaway}. In particular, for any $y_{\mathrm {min}}>10^{-39}$, the weakless universe with $v=M_{\rm Pl}$ is disfavored (this corresponds to the dashed vertical line in Fig.~\ref{fig:runaway}). Furthermore, the required minimal Yukawa values, which ameliorate the runaway behavior, are significantly smaller than the critical Yukawa value $\sim \Lambda_{\rm QCD}/M_{\rm Pl}$, which correspond to the light quark anthropic boundary, as denoted by the solid vertical line of Fig.~\ref{fig:runaway}.
Note that for $y_{\mathrm {min}}\sim  \Lambda_{\rm QCD}/M_{\rm Pl}$  the maximal value for the weak scale is as small as $\sim 10^9\,$GeV and the 
runaway towards the weakless universe is significantly ameliorated. 
In the white region of Fig.~\ref{fig:runaway} there are no light quarks in either type of 
universes. The price to pay in using this Yukawa distribution is that the resulting light
quarks will tend to be extremely light in general, as dictated  by the value of $y_{\mathrm {min}}$.

\section{Conclusions}

While it has been suggested that the multiverse may explain the
cosmological constant and the hierarchy problems, it is
generally believed that it cannot be of much use for the flavor
puzzle of the SM. The reason is that, while the properties of
our universe critically depend on the cosmological constant and
the Higgs VEV, many of the flavor parameters -- such as the
second and third generation masses and mixings -- do not seem
to affect the properties of ordinary matter. In this paper, we
have tried to challenge this point of view.

The starting observation is related to the recent discovery of
the Higgs boson. Given the known Higgs mass, a minor
modification of the top Yukawa can have dramatic consequences
for our universe, triggering an apocalyptic phase transition.
So, even if the value of a third-generation quark mass is
unrelated to the properties of matter, it can crucially
influence the evolution of our universe.

Motivated by this observation, we have studied the possibility
that the entire quark mass spectrum is determined by
statistical distributions of Yukawa couplings, under the
constraint of anthropic conditions. Our assumption is that the
SM Yukawa couplings originate from some high-scale flavor
dynamics, whose parameters scan within the multiverse.

Our goal is relatively modest. We do not aim at predicting
precisely the values of the quark masses and mixings, but we
are satisfied with showing that statistics could be the reason
behind the observed pattern. The only features of the quark
mass spectrum that we want to explain are: {\it (i)} two
ultra-light quarks ($u$, $d$) with masses much smaller than
$\Lambda_{\rm QCD}$; {\it (ii)} three quarks ($s$, $c$, $b$)
with Yukawas clustered around $\Lambda_{\rm QCD}/M_W$, the
``chemistry" anthropic boundary; {\it (iii)} one heavy quark
with Yukawa of order one, at the border of the ``stability"
anthropic limit.

For Yukawa distributions that are peaked towards small values 
when evaluated
in the UV, RG effects can generate in the IR a second peak for
large Yukawa couplings, as a consequence of a pseudo fixed
point. However, the values of the Yukawa corresponding to the
second peak are excluded by vacuum stability requirements, for
a Higgs mass of about 125--126 GeV. As a result, RG effects
cannot explain the near-criticality of the top mass, but they
help to increase the probability of obtaining  some heavy
quarks in the spectrum.

Especially interesting is the case of Yukawa distributions
emerging from strong dynamics at the high scale. In this case,
already in the UV there could be a tendency to populate
simultaneously the regions of light and heavy quarks, while
disfavoring the intermediate range. We find a plausible
probability that an average universe (which satisfies anthropic
conditions) has a pattern of quarks similar to what we observe,
with one heavy quark at the edge of the stability region.

\section*{Acknowledgments}

The authors are grateful to Oram Gedalia for many useful 
discussions and for actively collaborating in this work till its last 
stage. GP also thanks Lawrence Hall for discussions.  GP is the 
Shlomo and Michla Tomarin development chair,
supported by the grants from GIF, Gruber foundation, IRG, ISF and 
Minerva.

\end{document}